\documentclass{llncs}
\usepackage{graphicx}

\usepackage{amsthm}
\usepackage{multirow}
\usepackage{url}
\usepackage{hyperref}
\usepackage{calligra}
\usepackage{multirow}

\begin{document}

\title{Exposing Provenance Metadata Using Different RDF Models}
\author{Gang Fu\inst{1}, Evan Bolton\inst{1}, Nœria Queralt Rosinach\inst{2}, Laura I. Furlong\inst{2}, \\Vinh Nguyen\inst{3},  Amit Sheth\inst{3}, Olivier Bodenreider\inst{4}, \\Michel Dumontier\inst{5}}
\institute{\email{gang.fu@nih.gov}, \email{bolton@ncbi.nih.gov}\\ National Center of Biotechnology Information, National Institute of Health, \and \email{nqueralt,lfurlong@imim.es} \\Universitat Pompeu Fabra, Barcelona, Spain \and  \email{vinh,amit@knoesis.org}\\Kno.e.sis Center, Wright State University, Ohio, USA \and \email{olivier@nlm.nih.gov}\\National Library of Medicine, National Institute of Health, Maryland, USA \and \email{michel.dumontier@stanford.edu}\\Stanford Center for Biomedical Informatics, Stanford University, Stanford, USA}
\maketitle
\pagestyle{empty}

\begin{abstract}

A standard model for exposing structured provenance metadata of scientific assertions on the Semantic Web would increase interoperability, discoverability, reliability, as well as reproducibility for scientific discourse and evidence-based knowledge discovery. Several Resource Description Framework (RDF) models have been proposed to track provenance. However, provenance metadata may not only be verbose, but also significantly redundant. Therefore, an appropriate RDF provenance model should be efficient for publishing, querying, and reasoning over Linked Data. In the present work, we have collected millions of pairwise relations between chemicals, genes, and diseases from multiple data sources, and demonstrated the extent of redundancy of provenance information in the life science domain. We also evaluated the suitability of several RDF provenance models for this crowdsourced data set, including the N-ary model, the Singleton Property model, and the Nanopublication model. We examined query performance against three commonly used large RDF stores, including Virtuoso, Stardog, and Blazegraph. Our experiments demonstrate that query performance depends on both RDF store as well as the RDF provenance model. 

\end{abstract}

\section{Introduction}
Evidence and provenance are key aspects of a healthy scientific discourse. A standard model to provide structured and interoperable metadata linked to scientific assertions is of increasing interest \cite{sahoounified:2011,machado:2015}. The Resource Description Framework (RDF), the lingua franca for the Semantic Web, offers the building blocks by which statements can be provided along with their metadata. Structured metadata, such as whether the resource was manually curated or automatically text mined from scientific literature, is key to assessing quality of information. Hence, a scalable and well-designed RDF-based metadata model is crucial for knowledge integration.

Specifying the provenance of a single entity can be easily achieved using existing RDF terminologies such as PROV. However, it is the specification of the provenance of a binary or n-ary relation which remains non-standard. Several models for exposing the provenance metadata of the relations have been proposed including adding provenance annotations to i) an instance of a class that represents the n-ary relation (N-ary model) \cite{nary:w3}; ii) an instantiated property, i.e. Singleton property (SP) model \cite{Nguyen:2014:DLR:2566486.2567973}; and iii) a graph that contains the relational assertions, i.e. Nanopublication model \cite{groth2010anatomy}. In the life sciences, the N-ary model has been used to capture the provenance information for protein-protein interactions (i.e. iRefIndex database \cite{razick2008irefindex}) and text-mined gene-disease interactions (i.e. DisGeNET \cite{bauer2010disgenet}), while the recently proposed SP model \cite{Nguyen:2014:DLR:2566486.2567973} has been used across elements of biomedical and material sciences. Despite their use to represent various data, no study has yet been performed to examine the advantages and disadvantages of all these models using a common dataset.

In the present study, we aim to evaluate the consequence of using different RDF models to capture provenance metadata for life science data. We examine the number of triples generated and query performance on three RDF stores: Virtuoso \cite{virtuoso}, StarDog \cite{stardog}, and BlazeGraph \cite{blazegraph}. Regarding to the provenance metadata of the relational assertions, we consider the data source, the supporting scientific publication, and the biological species where the given assertion holds true. In addition to the three basic RDF models described above, we also examine the implementions of the so-called \textit{cardinal assertion model} that was first introduced by Nanopublications \cite{gibson2012towards} on the N-ary and SP models, to create a non-redundant network of assertions. This consideration is particularly important as there exists substantive overlap in the assertions from multiple databases. For instance, the asserted relation between dexamethasone (PubChem Compound 5743) and glucocorticoid receptor (GR) (NCBI Gene 2908) was mentioned by four different data sources, but each data source cites an entirely different set of scientific publications in support of the assertion. This work is crucial for the efficient implementation of scalable, interoperable, and extensible knowledge models for open data sources including PubChemRDF \cite{fu:2015}, Bio2RDF \cite{belleau2008bio2rdf}, and DisGeNET RDF\cite{pinero2015disgenet}.

\section{Methods}
\label{methods}
\subsection{Dataset preparation}
We generated a reference dataset of pairwise relations between chemicals, genes, and diseases from multiple data sources across life science domain. The chemical-disease relations were obtained from National Drug File Reference Terminology (NDFRT) \cite{brown2004va}, CTD \cite{davis2015comparative}, KEGG \cite{kanehisa2011kegg}, and SIDER \cite{kuhn2010side}; chemical-gene relations were obtained from CTD \cite{davis2015comparative}, DrugBank \cite{knox2011drugbank}, KEGG \cite{kanehisa2011kegg}, IUPHAR-DB \cite{sharman2012iuphar}, and ChEMBL \cite{gaulton2012chembl}; protein-protein relations were obtained from iRefIndex \cite{razick2008irefindex} and BioGRID \cite{stark2006biogrid}; gene-disease were contributed by DisGeNET \cite{bauer2010disgenet}. All chemicals were represented using PubChem Compound identifiers (CIDs), all genes were represented using National Center for Biotechnology Information (NCBI) Gene identifiers (GIDs), and all diseases were represented using the Unified Medical Language System (UMLS) Concept Unique Identifiers. The pairwise relations were normalized using the modified Semantic Network standard vocabulary \cite{rosemblat2013methodology}. The interrelations between biomedical entities (chemicals, genes, and diseases) constitute a semantic network, and SPARQL queries were used to explore the network topology on behalf of evidence-based hypothesis generation. However, it is fairly common to collect the identical assertion from multiple sources, in particular, for such a consolidated knowledge base. Hence, additional constraints were applied in the searching strategies.

\subsection{RDF model construction}

\begin{figure}[h!]
 \centering
 \includegraphics[width=0.95\textwidth]{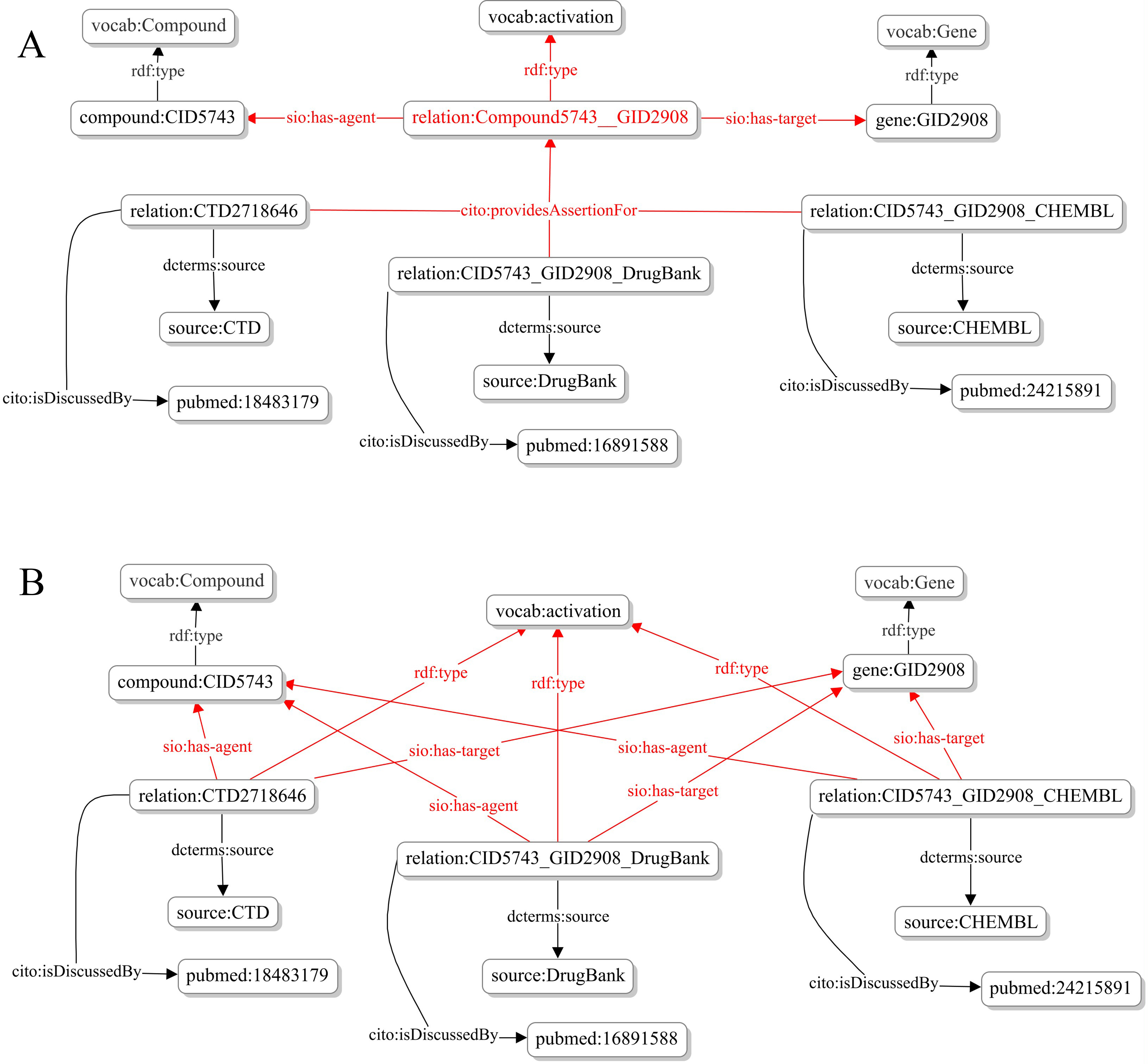}
 \caption{Graphical representation of N-ary model for the relation between compound CID5743 and gene 2908: (a) with cardinal assertion (Model I); (b)  without cardinal assertion (Model II). \label{fig1}}
\end{figure}

\begin{figure}[h!]
 \centering
 \includegraphics[width=0.95\textwidth]{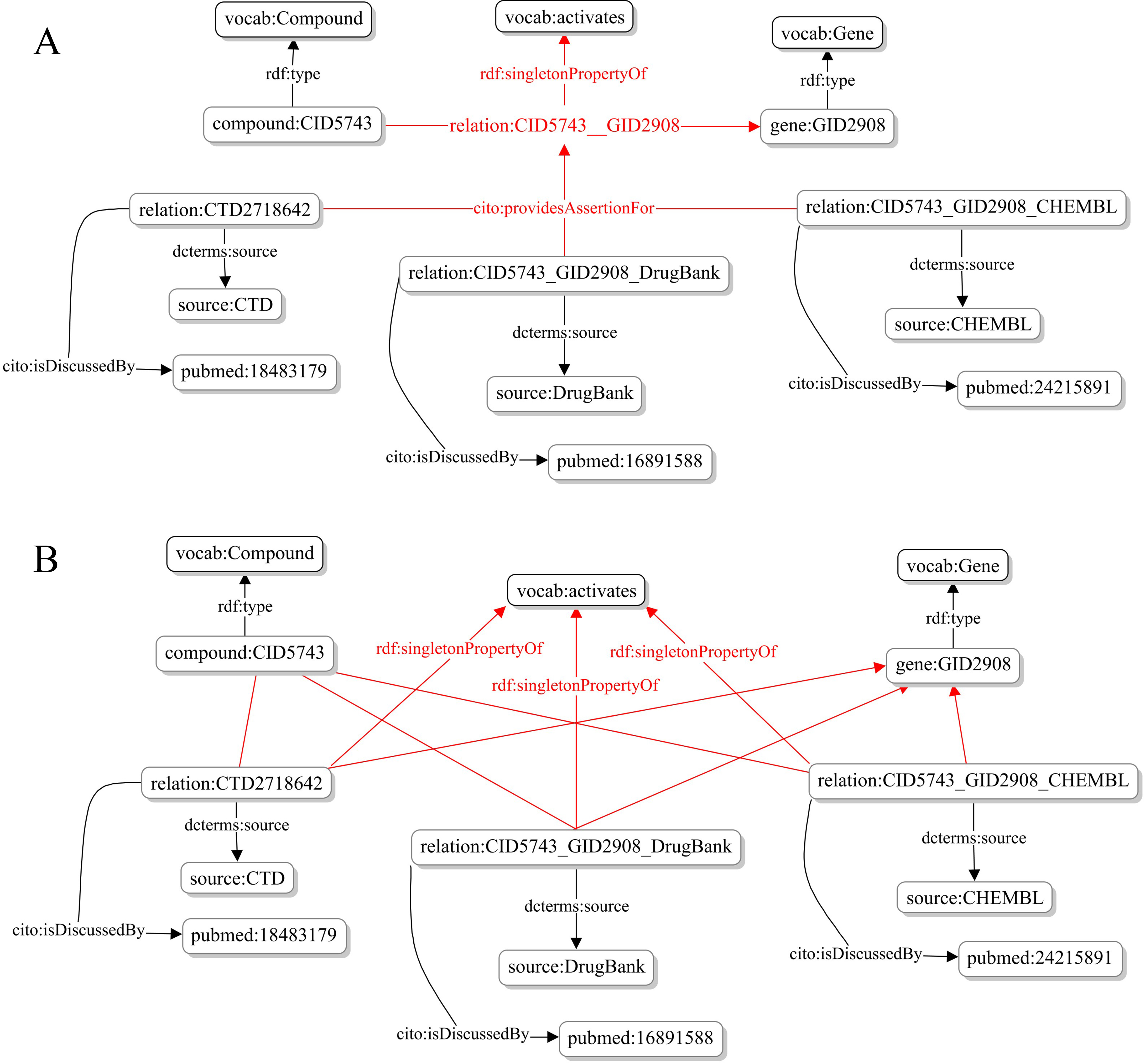}
 \caption{Graphical representation of SP model for the relation between compound CID5743 and gene 2908: (a) with cardinal assertion (Model III); (b) without cardinal assertion (Model IV). \label{fig2}}
\end{figure}

\begin{figure}[h!]
 \centering
 \includegraphics[width=0.95\textwidth]{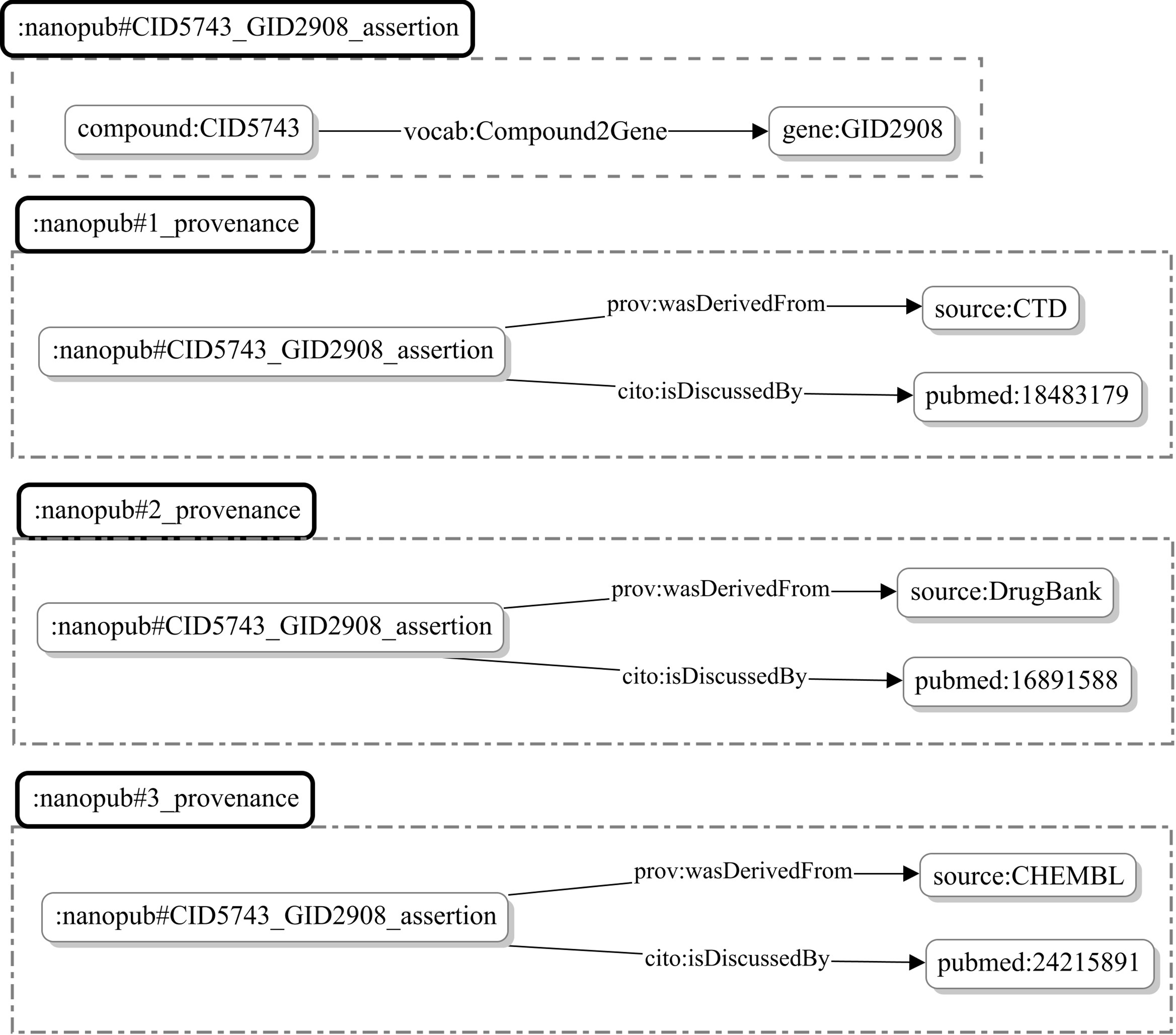}
 \caption{Graphical representation of Nanopublication model (Model V) for the relation between compound CID5743 and gene 2908. \label{fig3}}
\end{figure}

Five RDF models were studied, including N-ary model with and without cardinal assertion (Fig. \ref{fig1}), SP model with and without cardinal assertion (Fig. \ref{fig2}), and the Nanopublication model (Fig. \ref{fig3}). Only the assertion graphs and the provenance graphs were considered in the Nanopublication model. In both N-ary and SP cardinal assertion variants, a predicate \verb|cito:providesAssertionFor| is used to link the cardinal assertion of the pairwise relation to the multiple evidence (Fig. \ref{fig1}A and \ref{fig2}A). Without cardinal assertion, the pairwise relation would be asserted redundantly by multiple data sources (Fig. \ref{fig1}B and \ref{fig2}B). In the Nanopublication model variant A, one assertion graph may correspond with one or more than one provenance graphs (Fig. \ref{fig3}). In the following comparative analysis, Model \textbf{I} refers to the N-ary model with cardinal assertion, Model \textbf{II} refers to the N-ary model without cardinal assertion, Model \textbf{III} refers to the SP model with cardinal assertion, Model \textbf{IV} refers to the SP model without cardinal assertion, and Model \textbf{V} refers to the Nanopublication model.

\subsection{Query formulation}
An interesting research topic in drug discovery is to determine which proteins are responsible for eliciting particular drug side effects. We formulated SPARQL queries to examine this question using different levels of complexity (\textbf{Q1}, \textbf{Q2}) and provenance constraints (\textbf{Q3}, \textbf{Q4}). \textbf{Q1} explores the hypothesis that if chemical \textbf{A} inhibits gene \textbf{B}, and gene \textbf{B} interacts with gene \textbf{C}, and gene \textbf{C} is linked to disease \textbf{D}, then the above path can be used to explain the disease/adverse side effect \textbf{D} caused by chemical \textbf{A}. It should be noted that the observed side effect can be explained in several ways: either the aforementioned three-step indirect paths, or the two-step indirect path involving only the chemical-gene interaction and gene-disease associations. Therefore, we have constructed another query, i.e. \textbf{Q2}, to filter out the diseases that are associated with genes that directly interact with the given chemical. The first two queries do not take into account the provenance metadata, and it is usually the case that only the integrated assertions are considered on behalf of hypothesis generation and knowledge discovery. \textbf{Q3} narrows down the search results by applying data source constraints. \textbf{Q4} restricts by number of aggregated evidence on \textbf{Q1}: such that the query only considers the pairwise relations in the indirect path that have more than one supporting literature references. 

We carried out \textbf{Q1} through \textbf{Q4} on six chemicals that have extensive biomedical annotations from multiple data sources: propranolol (CID4946), clotrimazole (CID2812), mitoxantrone (CID4212), risperidone (CID5073), chlorpromazine (C-ID2726), and haloperidol (CID3559). There are hundreds of similar compounds in the integrated dataset and they are of key interest in the context of drug repurposing and development. 

	All queries were performed against three RDF stores without further tuning: open source Virtuoso 7.1, Stardog 2.2, and Blazegraph 1.5. The configuration allowed up to 16 GB memory for each RDF store to run queries, which were performed on cold cache. The Log10 transformations of the execution time in millisecond were illustrated in boxplot; the averages and standard deviations of the execution time in seconds were summarized as well in the comparative analysis.

The data sets and the SPARQL queries are available at:
\url{http://figshare.com/articles/Provenance_RDF_Models/1399197}.

\section{Results and Discussions}

\subsection{Data set statistics}

We first compared the total number of triples that each RDF model contains. The most efficient RDF model is SP model without cardinal assertion (Model \textbf{IV}), which contains 17,239,427 triples, and the cardinal assertion of SP model (Model \textbf{III}) increased the total number of triples by about 14\% to 19,575,298. For N-ary models, the cardinal assertion also increased the total number of triples by about 6\%, from 21,445,348 (Model \textbf{II})  to 22,787,218 (Model \textbf{I}). The N-ary model requires two triples (predicates \verb|sio:has-agent| and \verb|sio:has-target|) to represent the agent and target in a biological process, while the SP model maintains the previous binary relation structure in only one triple. Hence, with the cardinal assertion, the N-ary model (Model \textbf{I}) contains 3,211,920 more triples ($\sim$ 16\%) in comparison with the SP model (Model \textbf{III}), and without the cardinal assertion, the N-ary model (Model \textbf{II}) contains even more triples (4,205,921 triples) in contrast to the SP model (Model \textbf{IV}). The Nanopublication model is the most verbose model in this regard, which contains 27,605,782 triples distributed in 8,251,238 graphs. 

We also studied the amount of evidence associated with each relational assertion to illustrate the degree of redundancy with respect to the identical pairwise relations in the life science domain. We only examine object property instances representing the pairwise relations that were created in the SP models (Model \textbf{III} and Model \textbf{IV}), as the degree of redundancy is same across other RDF models. The total number of unique subjects in the SP models with and without cardinal assertion are 7,654,605 (Model \textbf{III}) and 4,442,685 (Model \textbf{IV}), respectively. The difference between the two numbers accounts for the total number of object property instances arbitrarily created for the cardinal assertions. If there are multiple cases of evidence for a given assertion, the cardinal assertion variant may reduce the total number of triples to express the same information, however, if there is only one case of evidence for a given assertion, the cardinal assertion will increase the total number of triples. Hence, whether the cardinal assertion can reduce the total number of triples depends on the extent of redundancy of the identical pairwise relations in the data set. Among 3,211,920 cardinal assertions, 2,800,124 ($\sim$87\%) of them are only associated with one evidence, 238,558 ($\sim$7\%) of them are associated with two cases of evidence, 67,088 ($\sim$2\%) of them are associated with three cases of evidence, and 98,625 ($\sim$3\%) of them are associated with more than three cases of evidence. The pairwise relations between PubChem compound CID5694 and NCBI gene GID5465 is associated with the most number of cases of evidence (3,096). Although there were many redundant assertions from multiple data sources, the majority have only one supporting evidence. Hence, the increase in the total number of triples were largely attributable to publication assertions. 

\subsection{Query performance evaluation}

We undertook a performance evaluation using three RDF databases (see Table \ref{table1}). With Virtuoso, the SP models with and without cardinal assertion (Model \textbf{III} and \textbf{IV}) largely outperformed the other models. \textbf{Q1} and \textbf{Q2} executed roughly 100 times faster on the SP models as compared to the N-ary models. Although Model \textbf{V} yielded comparable performance with Model \textbf{III} and I\textbf{V} in \textbf{Q1}, the additional filtering constraint made it much slower in \textbf{Q2}. In \textbf{Q4}, Model \textbf{III}, \textbf{IV}, and \textbf{V} performed similarly, which are 10 times faster than Model \textbf{II} and 100 times faster than Model \textbf{I}. In general, Virtuoso performed best using the SP models. With the Stardog RDF store, the N-ary models and the SP models were comparable in performance, but they always outperformed Nanopublication model. In particular, when the aggregated evidence was considered in \textbf{Q4}, both N-ary and SP models with and without cardinal assertion were carried out over 10 times faster than the Nanopublication model. Using Blazegraph, the Nanopublication model generally outperformed other models. In particular, \textbf{Q1} and \textbf{Q2} were carried out over 10 times faster in Model \textbf{V} rather than in other models.

Without querying the provenance metadata, the models with cardinal assertion (Model \textbf{I} and \textbf{III}) always yielded better performance in comparison with the models without cardinal assertion (Model \textbf{II} and \textbf{IV} accordingly). Hence, if we remove the redundant identical assertions from various data sources in both N-ary and SP models, the graph traversal-like queries can be executed much faster. If we think of conjunctive queries (i.e. graph traversal or inner join) as performing Cartesian products, the computational costs go up exponentially as the number of data items increase. Hence, the redundant pairwise relations cost much more time rather than cardinal assertions in \textbf{Q1} and \textbf{Q2}. However, if the provenance restrictions were considered, the model without cardinal assertion (Model \textbf{II} and \textbf{IV}) usually outperformed, except the \textbf{Q3} of the SP models executed in Stardog and \textbf{Q4} of both N-ary and SP models executed in Blazegraph. But the difference of query performance were usually small, except for the \textbf{Q4} of the N-ary models executed in Virtuoso, and the \textbf{Q3} of both N-ary and SP models executed in Blazegraph. So in general, if the provenance restrictions were considered, the models with and without cardinal assertion were comparable.

\begin{table}
\caption{The average execution time and standard deviation in seconds \label{table1}}
    \begin{tabular}{|c@{\hspace{0.5em}}|@{\hspace{0.5em}}c@{\hspace{0.5em}}|c@{\hspace{0.5em}}|c@{\hspace{0.5em}}|c@{\hspace{0.5em}}|c@{\hspace{0.5em}}|c@{\hspace{0.5em}}|} \hline\noalign{\smallskip}
       &  & \textbf{Model I}$^a$     & \textbf{Model II}$^a$    & \textbf{Model III}$^a$     & \textbf{Model IV}$^a$     & \textbf{Model V}$^a$      \\ \hline \noalign{\smallskip}
     \multirow{8}{*} {Virtuoso} &  \multirow{2}{*} {Q1} & 44.827     & 269.854    & \textbf{0.665}     & 2.398     & 1.283      \\
            &   & ($\pm$15.918)  & ($\pm$99.266)  & ($\pm$0.263)  & ($\pm$1.082)  & ($\pm$0.212)   \\  \noalign{\smallskip}
              & \multirow{2}{*} {Q2} & 260.337    & 369.635    & \textbf{0.535}     & 2.375     & 585.52     \\
              &   & ($\pm$253.588) & ($\pm$120.482) & ($\pm$0.301)  & ($\pm$1.083)  & ($\pm$193.382) \\  \noalign{\smallskip}
              & \multirow{2}{*} {Q3} & 4.04       & 3.069      & 3.075     & \textbf{2.248}     & 2.287      \\
              &   & ($\pm$0.294)   & ($\pm$0.161)   & ($\pm$0.243)  & ($\pm$0.718)  & ($\pm$0.049)   \\  \noalign{\smallskip}
              & \multirow{2}{*} {Q4} & 352.312    & 14.994     & 2.201     & \textbf{1.953}     & 2.531      \\
              &   & ($\pm$204.483) & ($\pm$11.587)  & ($\pm$0.028)  & ($\pm$0.331)  & ($\pm$0.054)   \\ \hline \noalign{\smallskip}
     \multirow{8}{*} {StarDog}    & \multirow{2}{*} {Q1} & \textbf{1.906}      & 5.499      & 3.354     & 16.805    & 21.291     \\
              &   & ($\pm$0.214)   & ($\pm$0.191)   & ($\pm$0.783)  & ($\pm$13.315) & ($\pm$0.621)   \\  \noalign{\smallskip}
              & \multirow{2}{*} {Q2} & \textbf{2.45}       & 6.366      & 4.072     & 18.398    & 184.96     \\
              &   & ($\pm$0.262)   & ($\pm$0.208)   & ($\pm$0.492)  & ($\pm$14.383) & ($\pm$90.820)  \\  \noalign{\smallskip}
              & \multirow{2}{*} {Q3} & 2.738      & \textbf{1.291}      & 3.834     & 16.463    & 27.537     \\
              &   & ($\pm$0.240)   & ($\pm$0.068)   & ($\pm$0.500)  & ($\pm$14.277) & ($\pm$1.602)   \\  \noalign{\smallskip}
              & \multirow{2}{*} {Q4} & 14.344     & 9.084      & 9.575     & \textbf{8.698}     & 45.959     \\
              &   & ($\pm$1.576)   & ($\pm$0.147)   & ($\pm$0.751)  & ($\pm$0.467)  & ($\pm$1.263)   \\ \hline \noalign{\smallskip}
     \multirow{8}{*} {Blazegraph} & \multirow{2}{*} {Q1} & 11.087     & 54.74      & 33.597    & 41.491    & \textbf{0.732}      \\
              &   & ($\pm$5.129)   & ($\pm$32.618)  & ($\pm$10.515) & ($\pm$11.862) & ($\pm$0.133)   \\  \noalign{\smallskip}
              & \multirow{2}{*} {Q2} & 10.853     & 56.599     & 33.469    & 41.099    & \textbf{4.215}      \\
              &   & ($\pm$4.494)   & ($\pm$34.590)  & ($\pm$10.292) & ($\pm$12.697) & ($\pm$1.635)   \\  \noalign{\smallskip}
              & \multirow{2}{*} {Q3} & 10.944     & 1.56       & 6.05      & \textbf{0.465}     & 0.581      \\
              &   & ($\pm$4.915)   & ($\pm$0.533)   & ($\pm$1.284)  & ($\pm$0.054)  & ($\pm$0.071)   \\  \noalign{\smallskip}
              & \multirow{2}{*} {Q4} & 83.384     & 117.131    & \textbf{74.505}    & 89.054    & 80.729     \\
              &   & ($\pm$1.612)   & ($\pm$2.721)   & ($\pm$1.570)  & ($\pm$1.119)  & ($\pm$1.436)   \\ \hline \noalign{\smallskip}
    \end{tabular}
\end{table}
$^a$ The average execution times are in the first line, and the standard deviations are in the second line within parenthesis; the best performance has been highlighted in bold.
\section{Conclusion}

In this study, we evaluated three existing RDF models and two cardinal assertion models for representing relations and exposing their provenance metadata. We examined the effect of each model on overall graph size and query time execution across three different RDF databases. Since our integrated life science dataset contained many duplicate assertions, graph traversal can be accomplished in a much more efficient way using the cardinal assertion. The redundant assertions add up a lot of computational overhead when searching through the integrated knowledge base for evidence-based hypothesis exploration. Surprisingly, we found that each RDF store performed the best using a different provenance model. It has been demonstrated that SPARQL queries may be executed in a RDF store specific manner in a previous analysis \cite{mironov2012gauging}. Our results drew a similar conclusion and may have contentious implications for the standardization of a provenance model, which should ideally be software/platform/system agnostic. A more extensive analysis with larger benchmark datasets and more query patterns would be helpful in the future study. 

\textbf{Acknowledgements}
This work was initiated at the 2014 BioHackathon in Fukashima; This research was supported [in part] by the Intramural Research Program of the National Library of Medicine; The research leading to these results has received support from Instituto de Salud Carlos III-Fondo Europeo de Desarrollo Regional (PI13/00082 and CP10/00524), the Innovative Medicines Initiative Joint Undertaking under grant agreements n¼ 115191 (Open PHACTS)], resources of which are composed of financial contribution from the European Union's Seventh Framework Programme (FP7/2007-2013) and EFPIA companiesÕ in kind contribution. The Research Programme on Biomedical Informatics (GRIB) is a node of the Spanish National Institute of Bioinformatics (INB).
\bibliographystyle{splncs03}
\bibliography{provenance}

\end{document}